\documentclass[twocolumn]{aastex62}

\def\Ha{\mathrm{H}\alpha}
\def\ff{f_\mathrm{f}}
\def\RP{R_\mathrm{P}}
\def\MP{M_\mathrm{P}}
\def\Mdot{\dot{M}}
\def\Lacc{L_\mathrm{acc}}
\def\LHa{L_\mathrm{\Ha}}
\def\FHa{F_\mathrm{\Ha}}
\def\Lsun{L_\odot}
\def\MJ{M_\mathrm{J}}
\def\RJ{R_\mathrm{J}}
\def\kms{\mathrm{km\,s^{-1}}}
\def\mcube{\mathrm{m^{-3}}}
\def\MJyr{\MJ\,\mathrm{yr^{-1}}}
\def\AHa{A'_\mathrm{H\alpha}}
\definecolor{mygreen}{rgb}{0.0,0.4,0.0}

\newcommand{\revise}{}
\newcommand{\reviseb}{}

\newcommand{\Prasun}{}
\newcommand{\revisec}{}

\usepackage{ulem}

\graphicspath{{./}{figures/}}

\begin{document}

\title{Constraining Planetary Gas Accretion Rate from H$\alpha$ Linewidth and Intensity: Case of PDS 70 b and c}

\correspondingauthor{Yuhiko Aoyama}
\email{yaoyama@tsinghua.edu.cn}

\author[0000-0003-0568-9225]{Yuhiko Aoyama}
\affiliation{ Department of Earth and Planetary Science, Graduate School of Science, The University of Tokyo, 7-3-1 Hongo, Bunkyo-ku, Tokyo 113-0033, Japan}
\affiliation{Institute for Advanced Study, Tsinghua University, Beijing 100084, People’s Republic of China}
\affiliation{Department of Astronomy, Tsinghua University, Beijing 100084, People’s Republic of China;}

\author[0000-0002-5658-5971]{Masahiro Ikoma}
\affiliation{ Department of Earth and Planetary Science, Graduate School of Science, The University of Tokyo, 7-3-1 Hongo, Bunkyo-ku, Tokyo 113-0033, Japan}
\affiliation{ Research Center for the Early Universe (RESCEU), Graduate School of Science, The University of Tokyo, 7-3-1 Hongo, Bunkyo-ku, Tokyo 113-0033, Japan}

%
%
\begin{abstract}
Recent observations of protoplanets embedded in \revise{circumstellar disks} have shed light on \revisec{the} planet formation process.
In particular, detection of hydrogen Balmer-line ($\Ha$) emission gives direct constraints \revisec{on} late-stage accretion \revisec{onto} gas giants.
Very recently \citet{Haffert+2019} measured the spectral line-widths, in addition to intensities, of $\Ha$ emission from the two protoplanets orbiting PDS~70.
Here, \Prasun{we study these protoplanets} \revisec{by} applying radiation-hydrodynamic models of the shock-heated accretion flow onto protoplanets that \citet{Aoyama+2018} has recently developed.
\Prasun{As a result,} we demonstrate that $\Ha$ line-widths combined with intensities lead to narrowing down the possible ranges of the protoplanetary accretion rate and/or mass significantly. 
While the current spectral resolution is not high enough to derive a definite conclusion regarding their accretion process, high-resolution spectral imaging of growing protoplanets is highly promising.
\end{abstract}

%
%
\section{Introduction}
Observation of growing protoplanets is a challenge, but provides crucial constraints \revisec{on} planet formation. 
\revise{Recent detections of $\Ha$ emission from young companions embedded in circumstellar gaseous disks 
sheds light on the late-stage gas accretion 
onto protoplanets
\citep{Close+2014,Sallum+2015,Wagner+2018}.}
Very recently, \citet{Haffert+2019} conducted follow-up observation with a high-resolution \revise{($R\sim2500$)} spectral imaging technique and thereby confirmed the previous detection of $\Ha$ emission from PDS~70~b. 
\revisec{They also} detected another source of strong $\Ha$ emission, which they identified as a second protoplanet in the PDS~70 system (i.e., PDS~70~c). 
Not only that, the high-resolution spectral imaging allowed them to obtain $\Ha$ line profiles for the two protoplanets.

Such $\Ha$ emission \Prasun{is likely to} \revisec{originate} from hot portions ($\sim 1 \times 10^4$~K) of infalling accretion flow \Prasun{onto} the protoplanet and \revise{circumplanetary disk}. 
\Prasun{There is} an empirical relation between the mass accretion rate and the full width at 10~\% of the maximum of \revisec{the} $\Ha$ line obtained for T Tauri star \revise{(TTS)} accretion \citep[e.g.][]{Hartmann+1994,Natta+2004}. 
\citet{Haffert+2019} estimated the mass accretion rates at $2 \times 10^{-8 \pm 0.4}$~$\MJ$~yr$^{-1}$ and $1 \times 10^{-8 \pm 0.4}$~$\MJ$~yr$^{-1}$ for PDS~70~b and c, respectively.
\Prasun{This} empirical relation \revise{is based} on the assumption that the accreting gas flowing along the stellar dipole magnetic field lines is hot enough to emit $\Ha$ radiation.

Gas accretion for protoplanets, however, may be different from stellar accretion. 
This is partly because \Prasun{of the fact that} the former is much less energetic than the latter due to their much lower mass and, \Prasun{hence}, shallower gravitational potential.
\revise{In the stellar case, 
high free-fall velocity causes a strong shock at the stellar surface, making the accreting gas hot enough to ionize hydrogen completely. This means that hydrogen line emission \Prasun{is not possible} \citep[][]{Hartmann+1994}. 
\Prasun{By} contrast, in the planetary case, the moderate shock heating makes a dominant contribution to $\Ha$ emission.}
To quantify the $\Ha$ emission from such shock-heated gas, 
\citet[][hereafter AIT18]{Aoyama+2018} developed a 1D radiation-hydrodynamic code that simulates non-equilibrium hydrogen-line emission from gas flow behind the accretion shock (see Section~\ref{sec: method}) and thereby demonstrated that the shock-heated gas generates significant hydrogen-line emission strong enough to be detected. 

In this \textit{Letter}, we report \Prasun{the study of}
the two accreting protoplanets in the PDS~70 system\Prasun{, applying the \revisec{models} of AIT18}. In particular, we demonstrate that the spectral \Prasun{line-width along with} \revise{its intensity (or the $\Ha$ luminosity)} as observational constraints \revisec{help} us \Prasun{to} narrow down the possible ranges of mass accretion rate onto the protoplanets.

%
%
\section{Model Description} \label{sec: method}
\subsection{Gas accretion feature}
\revise{
The geometry and flow pattern for protoplanetary accretion remain poorly understood. 
Some of the accreting gas falls almost freely onto the protoplanet directly, while some \revisec{settles} down \revisec{onto} the circumplanetary disk and \revisec{migrates} toward the central protoplanet. 
Even in the latter case, the accreting gas eventually falls freely from the inner edge of the circumplanetary disk onto the protoplanet, provided there is a wide gap between the protoplanet and circumplanetary disk (or an inner cavity) \citep{Konigl1991}. 
This is similar to the situation often considered and thus studied well in the case of the accretion of TTSs \citep[e.g., see the review of][]{Hartmann+2016}.
}

Both types of accretion flow would require a strong dipole magnetic field \revisec{for} the protoplanet. Its existence is predicted according to
the scaling law between the luminosity and magnetic-field strength of astronomical objects \citep{Christensen+2009}. 
Also, the predicted magnetic field of an accreting gas giant is strong enough to affect the accretion-flow pattern \citep{Batygin2018}.

\subsection{\revise{Modeling of H$\alpha$ emission}}
A common feature in both cases is that the gas flow collides \revisec{with} the protoplanet's surface \revisec{at the} a free-fall velocity.
Since the free-fall velocity is higher than the local sound one, the gas flow passes through shockwaves \revise{there.} 
\revisec{Shock} compression heats the gas to a temperature \revisec{on} the order of $10^4$~K, which is high enough to dissociate hydrogen molecules and ionize \revise{some of the} hydrogen atoms, producing free electrons.
The electrons collide with and excite the hydrogen atoms. 
De-excitation of the excited hydrogen results in line emission and cooling.

With the numerical code developed by \citet{Aoyama+2018}, 
we simulate the radiation hydrodynamics of the 1D gas flow behind the shock front, 
including calculations of chemical reactions, excitation/de-excitation of hydrogen atoms, and radiative transfer.
\revise{The collisional- and radiative-transitions of energy levels in hydrogen atoms are calculated in a time-dependent way with the transition rate coefficients from \citet{Vriens+Smeets1980}.
As demonstrated in \citet{Aoyama+2018}, since the shock-heated gas cools immediately and the H$\alpha$ emission occurs only in a thin layer below the shock front, the plane-parallel (or 1D) approximation is valid. 
} 
We assume an ideal gas mixture with the solar elemental abundances from \citet{Allen3rd}. 
The input parameters are the preshock velocity $v_0$ and the number density of hydrogen nuclei $n_0$.
The details of the AIT18 model are given in Section~2 of \citet{Aoyama+2018}. 
%

%
%
\section{Theoretical Emission Property}

\subsection{Spectral line-width}
\label{sec:width}

\begin{figure}
\plotone{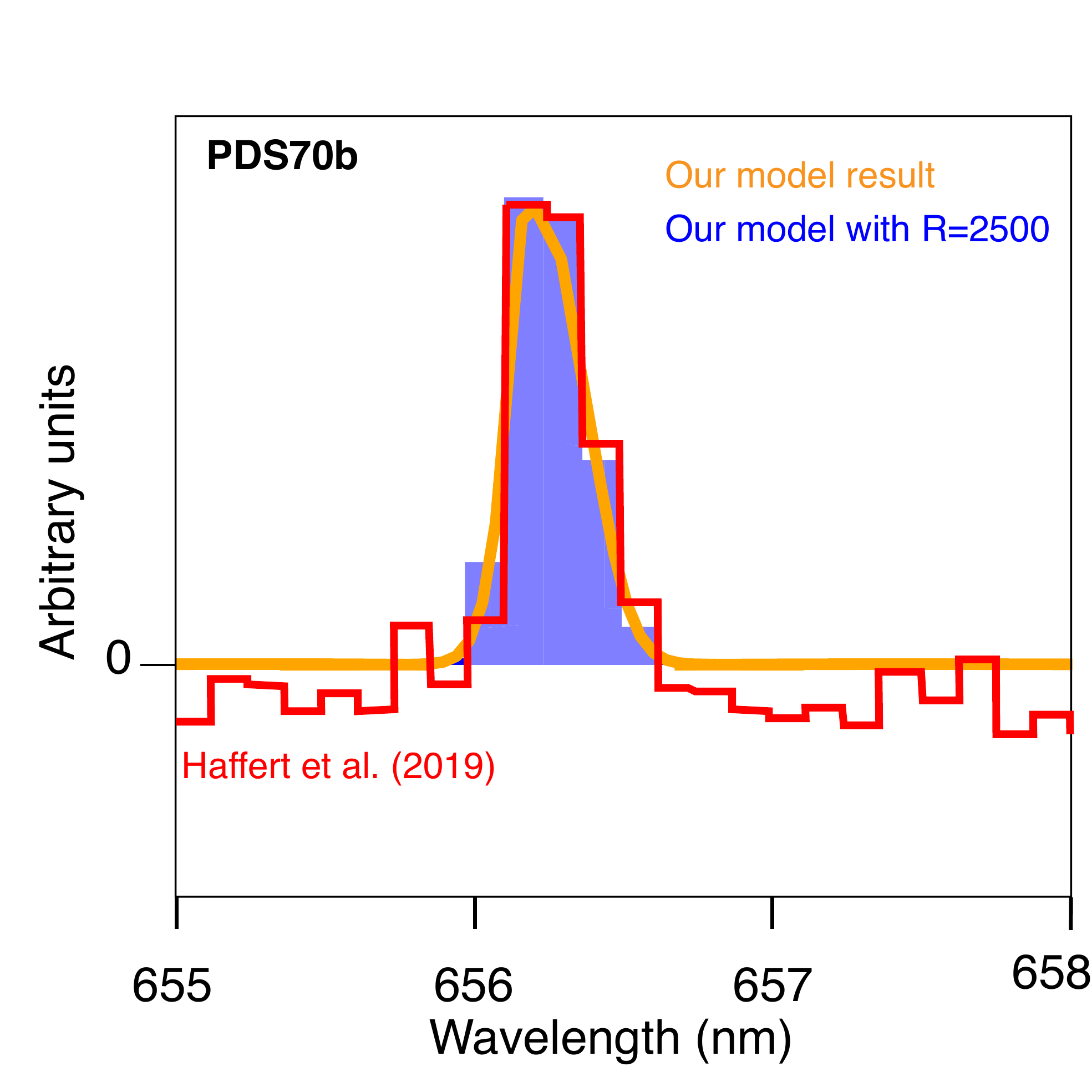}
\plotone{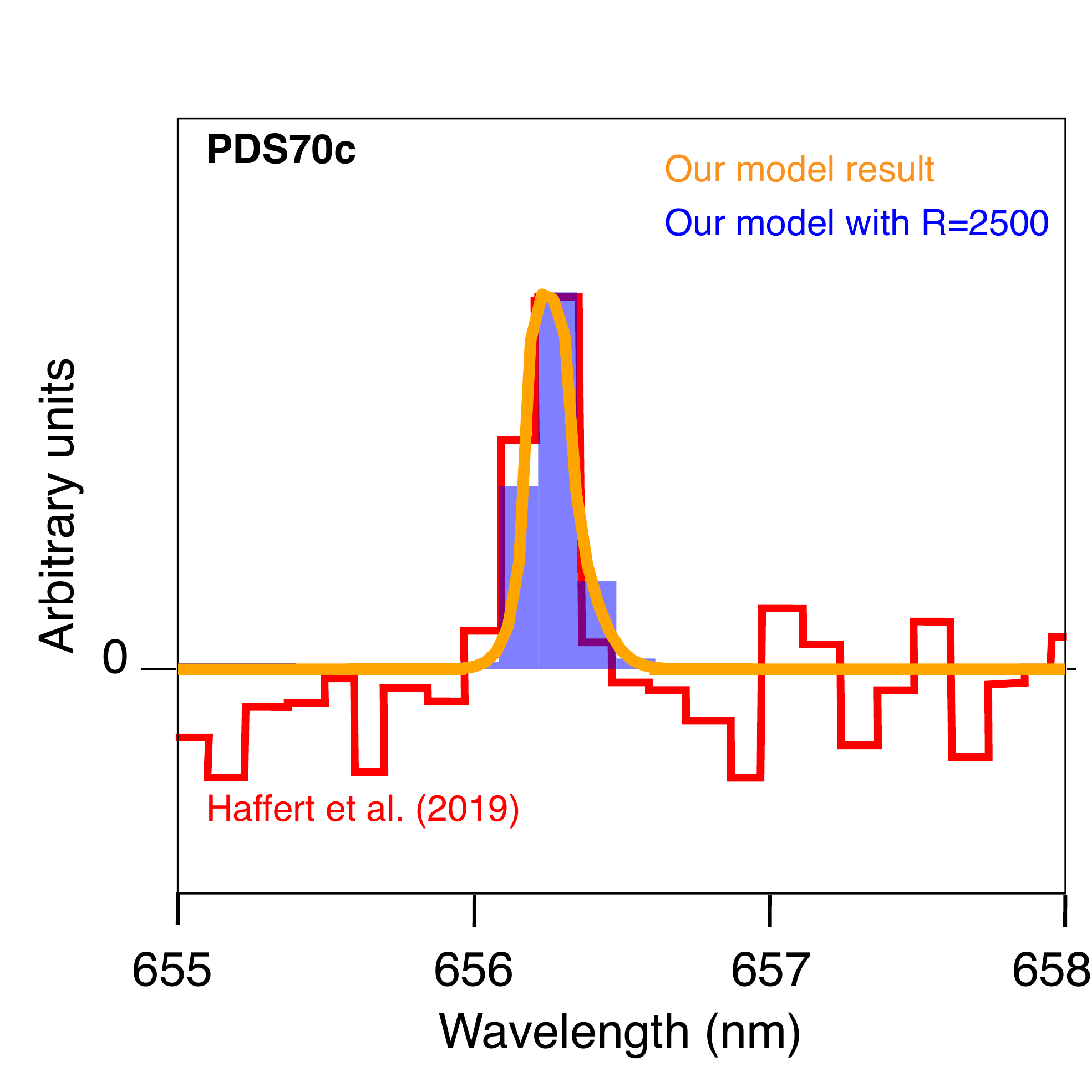}
\caption{
\revise{Simulated and observed} spectral emission profiles around the $\Ha$ line for PDS~70~b ($upper$ $panel$) and c ($lower$ $panel$). 
\revise{The observed profiles, which were} obtained with MUSE/VLT by \citet{Haffert+2019}, \revise{are} indicated by red lines.
As for \revise{the simulated} profiles, the raw data are shown \revisec{with} orange lines, while the data smoothed with a filter of $R$ = 2500 are shown \revisec{with} blue bars. 
Note that the quantities \revise{of} the vertical axis differ between the observed and calculated data; the observed data are the signal-to-noise ratio, whereas the raw and smoothed calculation data are the energy flux per unit wavelength and just the energy flux, respectively. 
Also, the calculated profiles have been artificially shifted by $-2.5$\,\AA\ and $-2.2$\,\AA\ for PDS~70~b and c, respectively, so as to coincide with the observed ones.
}
\label{fig:profile}
\end{figure}

First, we demonstrate that our 1D radiation-hydrodynamic models yield $\Ha$ line profiles that are consistent with \revisec{the} ones for PDS~70~b and c observed with MUSE/VLT by \citet[][]{Haffert+2019}.
Figure~\ref{fig:profile} compares the observed spectral emission profiles \revise{with simulated ones. The assumed 
values of the preshock velocity $v_0$ and the number density of hydrogen nuclei $n_0$} 
in our shock model are $v_0$ = 150~$\mathrm{km\,s^{-1}}$ and $n_0$ = $1 \times 10^{20}\,\mathrm{m^{-3}}$ for PDS~70~b (\textit{upper panel}) and $v_0$ = \revise{130} $\mathrm{km\,s^{-1}}$ and $n_0$ = $1 \times 10^{19}\,\mathrm{m^{-3}}$ for PDS~70~c (\textit{lower panel}).
(Note that these sets of $v_0$ and $n_0$ yield the full widths at 10\% and 50\% of the maximum of the protoplanet's H$\alpha$ emission lines that are consistent with the observed \revisec{values} for each planet; see \S~\ref{sec:constraint} for the details).
In the upper and lower panels, the red lines indicate the observed signal-to-noise ratio.
As for the model profiles, the orange line indicates the raw \revisec{line} energy flux per unit wavelength, while the blue one represents the energy flux smoothed with a filter of $R$ = 2500.
It does not matter that the quantities on the vertical axis of each panel differ between the synthesized and observed profiles,  
since the focus is on the spectral profile rather than intensity here.
Also, though the wavelength is also arbitrarily shifted, it does not affect the following descriptions.

\begin{figure*}
    \centering
    \plottwo{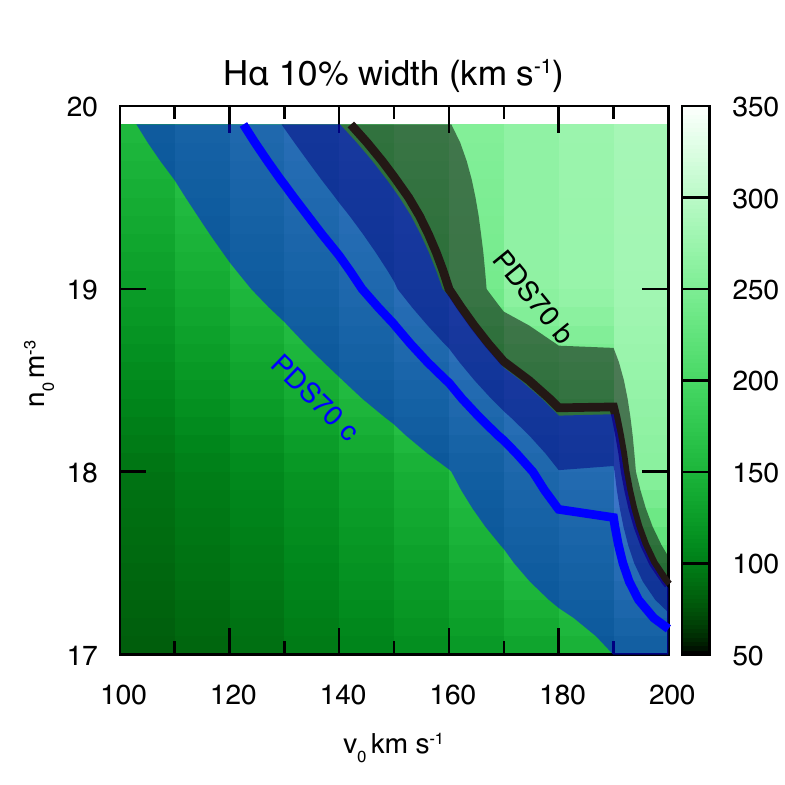}{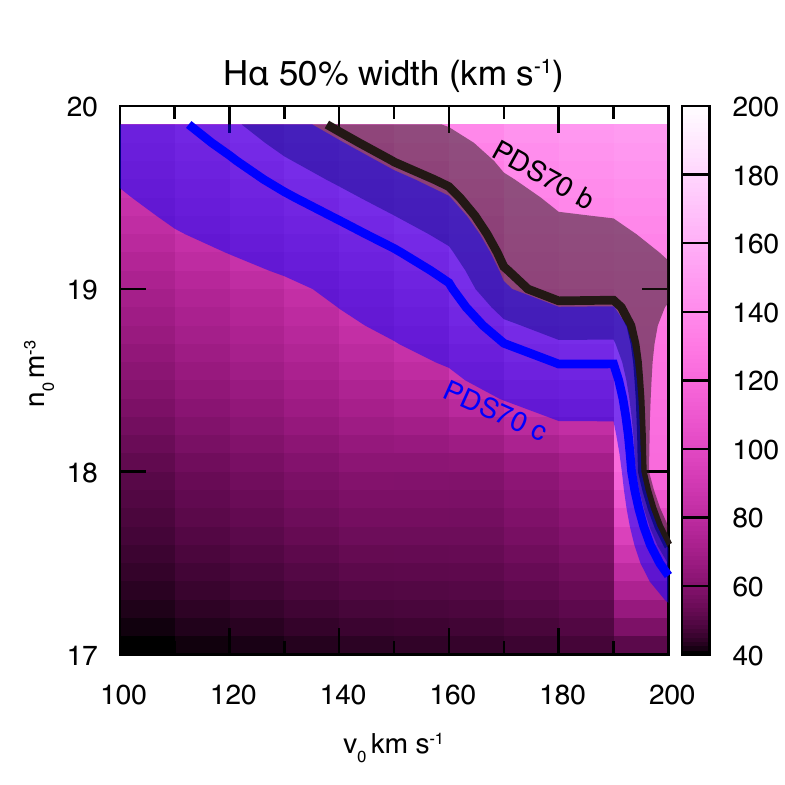}
    \caption{
    The full width at 10~\% (\textit{left panel}) and 50~\% (\textit{right panel}) of the maximum of the protoplanet's $\Ha$ emission line.
    Contour plots of the full width are shown as functions of the pre-shock velocity $v_0$ and number density $n_0$. 
    In each panel, the maximum likelihood values and 1$\sigma$ errors of the observed widths from \citet{Haffert+2019} are represented by thick lines and shaded areas, respectively; \revise{black} and \revise{blue} are for PDS~70~b and c, respectively.  
    }
    \label{fig:width}
\end{figure*}

For more quantitative comparison, rather than the shape of spectral profiles, we \revisec{also} focus on the spectral line-width. 
\citet{Haffert+2019} estimated the full widths at 10~\% and 50~\% of the maximum of the protoplanet's $\Ha$ emission line (simply the $\Ha$ 10\% and 50\% full widths, respectively, hereafter),
values that are listed in Table~\ref{tab:pds}. 

Figure~\ref{fig:width} shows the calculated $\Ha$ 10\% full width (\textit{left panel}) and 50\% full width (\textit{right panel}) as a function of the pre-shock velocity $v_0$ and number density $n_0$. 
\revisec{In} both panels, the $\Ha$ full width increases with pre-shock velocity. This is because a flow with higher velocity passes through a stronger shock and then becomes hotter, enhancing Doppler broadening, which is responsible for the width of the $\Ha$ line profile.

Secondly, it turns out that an increase in the number density results in increasing the $\Ha$ full width. 
This is not due to pressure broadening, which is negligibly small relative to Doppler and natural broadening, but due to the effect of absorption. 
The $\Ha$ radiation, which comes out deep below the shock front, propagates upward through the shock-heated gas toward the shock front. 
During \revisec{that} propagation, \revisec{some} of the $\Ha$ radiation is absorbed \citep{Aoyama+2018}.
The higher the gas density, the stronger the absorption of $\Ha$ radiation near the line center. 
Such a decrease in the line-peak intensity results in \reviseb{an increase} in the $\Ha$ full width because the latter is measured from the former.

\reviseb{In Figure~\ref{fig:width},} the observational results from \citet{Haffert+2019} are indicated by the black line for PDS~70~b and by the \revise{blue} line for PDS~70~c with 1$\sigma$ errors indicated by the same-color shades. 
From this figure, we obtain possible ranges of $v_0$ and $n_0$ \revisec{for} the accretion flow toward the two protoplanets (see Section~\ref{sec:constraint} for further discussion).

\subsection{$\Ha$ luminosity}
\label{sec:luminosity}
\begin{figure}
    \centering
    \epsscale{1.25}
    \plotone{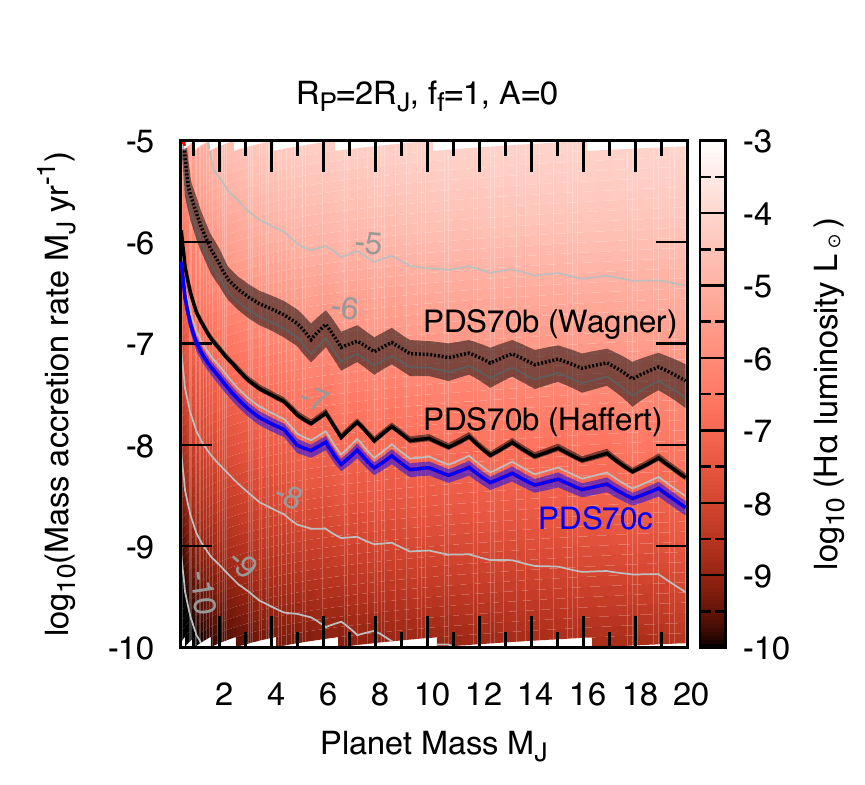}
    \caption{
    Color contour plot of the protoplanet's $\Ha$ luminosity as a function of protoplanet mass and mass accretion rate. 
    As indicated in the figure, the lines and shaded areas represent the observed maximum likelihood values and 1$\sigma$ errors of $\Ha$ luminosity for 
    PDS~70~b, $1.4\pm0.6\times10^{-6}\Lsun$ \citep[\textit{black dashed line};][]{Wagner+2018} and 
    $1.6\pm 0.14 \times 10^{-7}\Lsun$ \citep[\textit{black solid line};][]{Haffert+2019}, and 
    PDS~70~c, $7.6\pm1.3\times10^{-8}\Lsun$ \citep[\textit{blue line};][]{Haffert+2019}. The gray thin lines are contour lines for $\Ha$ \revisec{luminosities} of $10^{-5}$ to $10^{-10}\Lsun$.
    Here, we have assumed the protoplanet radius $\RP=2\RJ$, the filling factor $\ff=1$, and the extinction $\AHa=0$.
    }
    \label{fig:luminosity}
\end{figure}
 
The values of the observed $\Ha$ luminosity for the two protoplanets reported by \citet{Wagner+2018} and \citet{Haffert+2019} are listed in Table~\ref{tab:pds}. In our model, we assume that a strong shock occurs at the surface of the protoplanet so that the H$\alpha$ luminosity is given by 
\begin{equation}
    \LHa= 4 \pi \RP^2 \ff \FHa 10^{-\reviseb{\AHa}},
    \label{eq:luminosity}
\end{equation}
where $\RP$ is the planetary radius, 
$\ff$ is the fractional area of planetary surface where the accretion flow eventually emits $\Ha$ radiation, which is termed the filling factor, and
$\FHa$ is the H$\alpha$ energy flux per unit area.
The factor \reviseb{$10^{-\AHa}$} represents extinction of the $\Ha$ radiation on the way from the protoplanet's emission surface to the observer (in addition to interstellar absorption). 
\revise{
The extinction is caused by the circumstellar disk, \revisec{disk} wind above the disk, accretion flow towards the protoplanet, and so on.
The exact values of $\ff$ and $\AHa$ are poorly known.}

Figure~\ref{fig:luminosity} shows the calculated value of $\LHa$ as a function of the protoplanet mass $\MP$ and the planetary mass accretion rate $\Mdot$; 
the latter is given by
\begin{equation}
\Mdot = 4\pi \RP^2 \ff \, \mu^\prime n_0 v_0 ,
\label{eq: mass accretion rate}
\end{equation}
where $\mu^\prime$ is the mean weight per hydrogen nucleus. 
Also, since the pre-shock velocity is assumed to be the free-fall one, the planetary mass is related to $v_0$ as
\begin{equation}
    \MP = \frac{\RP v_0^2}{2G},
\label{eq: free fall velocity}
\end{equation}
where $G$ is the gravitational constant.
Here we have assumed $\RP$ = 2~$\RJ$ and $\ff$ = 1. 
In this figure, the $\Ha$ luminosity is found to be almost proportional to protoplanet mass and mass accretion rate (i.e., $\LHa \propto \Mdot \MP$) 
for the following reason: From Eqs.~(\ref{eq:luminosity})--(\ref{eq: free fall velocity}),
\begin{equation}
    \label{eq:L_MdM}
    \LHa = \Mdot \frac{2G\MP}{\RP} \frac{\FHa(v_0,\,n_0)}{\mu'n_0 v_0^3} 10^{-\AHa}.
\end{equation}
\citet{Aoyama+2018} found $\FHa$ is roughly proportional to $n_0 v_0^3$.
Note that although absent in Eq.~(\ref{eq:L_MdM}), the filling factor $\ff$ affects $\LHa$ somewhat, because $\FHa/(n_0 v_0^3)$ is roughly constant but varies with $n_0$ due to optical-depth effects, and $n_0$ depends on $\ff$ as shown in Eq.~(\ref{eq: mass accretion rate}). 
However, $\ff$ is still less important than the other parameters for $\LHa$.

In Fig.~\ref{fig:luminosity}, like in Fig.~\ref{fig:width}, we indicate the observational results from \citet[][]{Wagner+2018} and \citet[][]{Haffert+2019}.
Infra-red observations estimate the masses of PDS~70~b and c at 2-17~$\MJ$ \citep[][]{Muller+2018} and 4-12~$\MJ$ \citep[][]{Haffert+2019}, respectively. 
According to Fig.~\ref{fig:luminosity}, in those ranges of protoplanet mass, the mass accretion rate for PDS~70~b is $\sim 1 \times 10^{-7}$~$\MJyr$ for the data from \citet{Wagner+2018} and $\sim 1 \times 10^{-8}$~$\MJyr$ for the data from \citet{Haffert+2019}, while that for PDS~70~c is $\sim 1 \times 10^{-8}$~$\MJyr$.
We discuss the difference in estimated mass accretion between the present and previous studies in Section~\ref{sec: previous study}.

Note that we have assumed that all of the accreting gas falls onto the surface of the protoplanet, \revisec{as in} the above studies \citep[][]{Wagner+2018,Haffert+2019}. 
Since the focus of this study is on the effect of \revisec{the} accretion shock on the $\Ha$ emission, detailed treatment of accretion flow toward the protoplanetary system including the \revise{circumplanetary disk} is beyond the scope of this study. A further detailed investigation is done in our forthcoming paper \citep{AGCI2019}.

%
%
\section{Discussion}
\subsection{Gas accretion rates for PDS~70~b and c} \label{sec:constraint}
\begin{figure*}
    \centering
    \plottwo{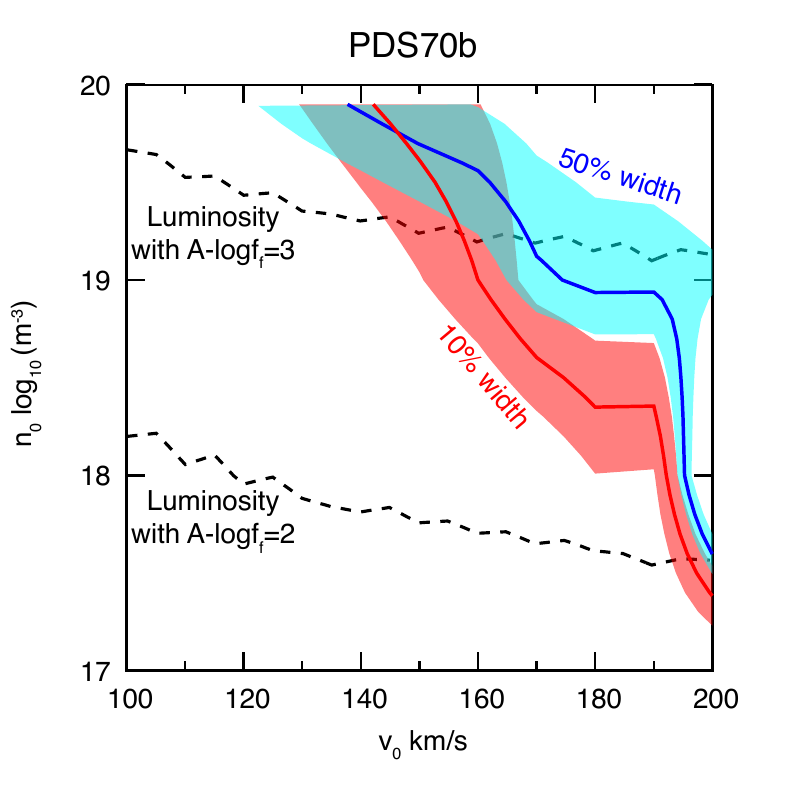}{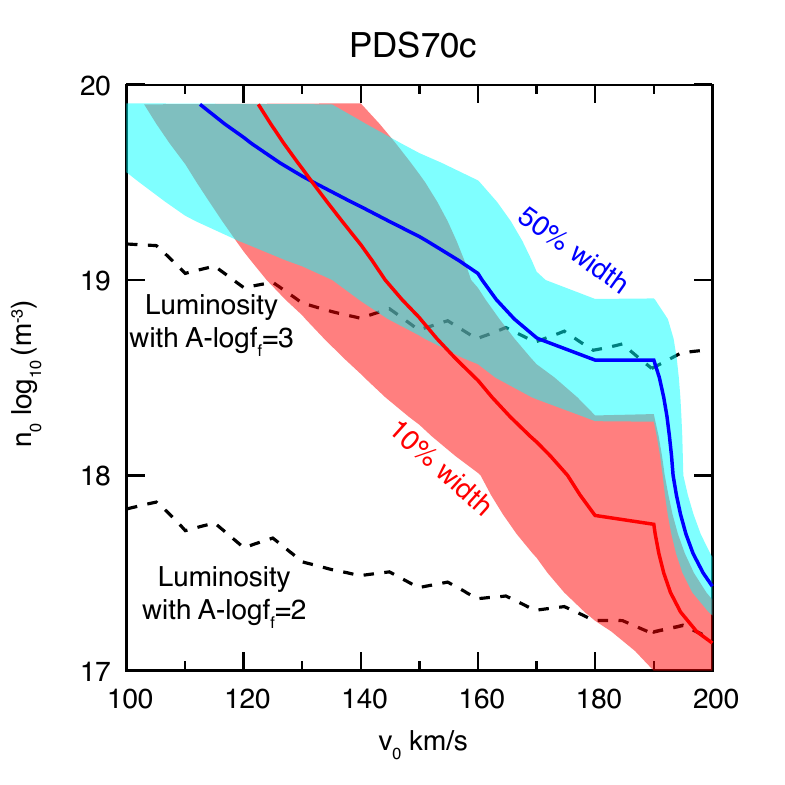}
    \caption{
    Properties of accretion shock on the protoplanetary surface inferred with three different observational \revise{constraints including the $\Ha$ luminosity, 10~\%  full width and 50~\% full width}. 
    The \revise{red} and \revise{blue} lines correspond to the maximum likelihood values of the $\Ha$ 10\,\% width and 50\,\% width, respectively (see also Fig.~\ref{fig:width}). The \revise{black} dotted lines represent the observed $\Ha$ luminosity for two choices of the uncertain parameter $\ff 10^{-A}$ (see Eq.~[\ref{eq:luminosity}]), namely $\AHa-\log\ff=2$ and $3$.
    }
    \label{fig:contour}
\end{figure*}

\begin{deluxetable}{l|l l}
\tablecaption{\revise{Properties of PDS70b and c}}
\tablehead{ & \colhead{PDS~70~b} & \colhead{PDS~70~c} }
\startdata
\multicolumn{3}{c}{\revise{H$\alpha$ observation}}\\ \hline
 10\% full-width [$\mathrm{km\,s^{-1}}$] & $224\pm24^{[1]}$ &$186\pm35^{[1]}$\\
 50\% full-width [$\mathrm{km\,s^{-1}}$] & $123\pm13^{[1]}$ & $102\pm19^{[1]}$\\
 Luminosity [$10^{-7}\Lsun$] & $1.6\pm0.14^{[1]}$ & $0.76\pm0.13^{[1]}$\\
 &$14\pm6^{[2]}$& non-detection$^{[2]}$ \\ \hline
\multicolumn{3}{c}{\revise{Estimated properties in this study$^\mathrm{[3]}$}} \\ \hline
 $\MP$ [$\MJ$] & $12$ & $10$ \\
 $\Mdot$ [$10^{-8}\MJ \, \mathrm{yr}^{-1}$]& $4$ & $1$ \\ 
 $\ff$ [$10^{-3}$]& 0.50 & 0.33 \\ \hline
 \multicolumn{3}{c}{\revise{Estimated properties in previous studies}} \\ \hline
 $\MP$ [$\MJ$] & $2$--$17$ & $4$--$12^\mathrm{[4]}$ \\
 $\Mdot$ [$10^{-8}\MJ \, \mathrm{yr}^{-1}$]& $2^\mathrm{[5][6]}$ & $1^\mathrm{[5][6]}$
\enddata
\tablecomments{[1]\citet{Haffert+2019}, [2]\citet{Wagner+2018}}, \revise{[3]\S\ref{sec:constraint}, [4]\citet{Muller+2018}, [5]\citet{Haffert+2019}, [6]\citet{Natta+2004}}
\label{tab:pds}
\end{deluxetable}

Combining the three kinds of observational datum, the $\Ha$ luminosity, 10\% full width, and 50\% full width, we narrow the possible ranges of the mass accretion rate and protoplanet mass for the two accreting protoplanets. In Fig.~\ref{fig:contour}, we show again the $v_0$-$n_0$ relationships for PDS~70~b and c derived from our models with \revisec{all} three observational constraints.
Since the theoretical estimate of $\Ha$ luminosity depends on two uncertain parameters\revisec{, namely} $\ff$ and $\AHa$ \revise{(see Eq.~[\ref{eq:luminosity}])}, we show the $v_0$-$n_0$ relationships for two different choices of $10^{\AHa} \ff^{-1}$ or $\AHa - \log \ff$ \reviseb{= 2 and 3}.
While correct values of $\AHa$ and $\ff$ are poorly understood for the case of planetary accretion, we note that their ranges are constrained in
the context of accreting TTS: Theoretical modeling of accretion \revisec{shocks} with $\ff$ = $\sim 10^{-5}$--$10^{-1}$ reproduces observed UV excesses of TTSs \citep[e.g. Table 12 of][]{Herczeg+Hillenbrand2008}.
Also, \citet{Wagner+2018} assumed $0\lesssim \AHa \lesssim 3$ as a likely range of $\AHa$.

The intersection point of the two lines of the maximum likelihood values \revisec{for} 10\% and 50\% full widths is $(v_0,\,n_0)$ = $(145\,\mathrm{\kms},\,6\times10^{19}\,\mcube)$ for PDS~70~b and $(130\,\kms, 3\times10^{19}\,\mcube)$ for PDS~70~c.
To reproduce the observed $\Ha$ luminosity, the value of $10^{\AHa}\ff^{-1}$ comes out to be $2\times10^{3}$ and $3\times 10^{3}$ for PDS~70~b and c, respectively. 

\revise{
To convert ($v_0$, $n_0$) to ($\MP$, $\Mdot$), we assume $\AHa=0$ and $\RP=2\RJ$.}
\revise{Then, substituting the values of $v_0$, $n_0$, and $\ff$ estimated above into Eqs.~(\ref{eq: mass accretion rate}) and (\ref{eq: free fall velocity}), we obtain $(\MP,\,\Mdot)$}
as $(12\MJ,\,4\times10^{-8}\MJyr)$ and $(10\MJ,\,1\times10^{-8}\MJyr)$ for PDS~70~b and c, respectively. \revise{The estimated values are listed in Table~\ref{tab:pds}.}
The assumption of $\AHa=0$ \revisec{is} a reasonable one, given \revisec{that} the gas falling onto the protoplanet is depleted of dust and thus optically thin in \revisec{the} late stages of planet formation.
Also, the resultant value of $\ff\sim10^{-3}$ is similar to that for the case of accretion shocks for low-mass stars \citep{Herczeg+Hillenbrand2008}. 
However, given the importance of the filling factor, detailed numerical simulations of protoplanetary gas accretion should be done \revisec{to determine} the exact value of $\ff$.

\subsection{Comparison with previous models \label{sec: previous study}}
\citet{Wagner+2018} and \citet{Haffert+2019} also estimated the mass accretion rate onto PDS~70~b and/or c, based on empirical relationships derived from observations of accreting, low-mass T Tauri stars.
Here, we discuss the difference between our model and the previous models, which yield different estimates of the mass accretion rates $\Mdot$.

To estimate $\Mdot$ from the observed $\Ha$ luminosity $\LHa$ for PDS~70~b,
\citet{Wagner+2018} used the empirical formula derived by \citet{Rigliaco+2012} which gives a relationship between $\LHa$ and the continuum integrated over all wavelengths $\Lacc$. They assumed that the latter (i.e., $\Lacc$) was related directly to the mass accretion rate.
For $\AHa =0$, for example, their estimated $\Mdot$ is $10^{-8.7\pm0.3}\MJyr$.
This value is smaller by one to two orders of magnitude than our estimate
(see black dotted line in Fig.~\ref{fig:luminosity}). 
\revisec{This} difference comes from the fact that hydrogen line emission accounts for a significant fraction of the total emission for the protoplanetary case, in contrast to the case of T Tauri stars \revise{(see \S~\ref{sec:tts} for the detail)}.
Relationships between $\LHa$ and $\Lacc$ and between $\LHa$ and $\Mdot$ applicable to planetary accretion are presented in our forthcoming paper \citep{AGCI2019}.

\citet{Haffert+2019} estimated the mass accretion rate to be $2\times 10^{-8\pm0.4}\,\MJyr$ and $1\times 10^{-8\pm0.4}\,\MJyr$ for PDS~70~b and c, respectively, from the empirical relationship between the mass accretion rate and the $\Ha$ 10\,\% full width derived by \citet{Natta+2004}. 
These values are, by chance, similar to our estimates given in Section~\ref{sec:constraint}.
The process of $\Ha$ emission, however, definitely differs between their and our models:
In contrast to our model, \citet{Natta+2004} considered that pre-shock gas flowing towards T Tauri stars (not shock heated gas) \revisec{is} hot enough to generate $\Ha$ emission. 
\revisec{For} different choices of $\ff$ and $\AHa$, the estimated value of $\Mdot$ differs considerably between \revisec{the} models.
Combined effects of post-shock emission \revisec{the} pre-shock emission are of great interest. Therefore it \revisec{will} be important to \revisec{conduct} detailed investigation to confirm whether enough heating occurs in the accretion flow for protoplanets.

\subsection{Difference between planetary and stellar cases}
\label{sec:tts}
\revise{
The key difference between planetary and stellar accretion is in the free-fall velocity, shock strength. \revisec{This} determines whether the shock-heated gas becomes hot enough to ionize the hydrogen completely; of course, completely ionized hydrogen never emits $\Ha$ radiation.
Stellar accretion \revisec{shocks are} strong enough in that sense.
Therefore, previous studies of stellar accretion focus on \textit{pre}-shock gas as the source of $\Ha$ emission. 
In contrast, in the case of planetary accretion,
the shock-heated gas becomes hot enough to excite but not enough to ionize hydrogen.
Thus, the \textit{post}-shock gas is the main source of $\Ha$ emission. \revisec{Pre-shock} heating is unlikely for planetary accretion.
Therefore, we have applied the post-shock emission model \citep{Aoyama+2018} to the planetary-mass objects PDS~70~b and c in this study.
}

\section{Concluding Remarks}
\label{sec:future}
In this study, we have applied our planetary accretion shock model \citep[][]{Aoyama+2018} to the two accreting proto-gas giants\Prasun{-}PDS~70~b and c, for which $\Ha$ emission fluxes and profiles were very recently observed \citep[][]{Wagner+2018,Haffert+2019}. 
We have demonstrated that the new observational data, namely, spectral profiles, combined with $\Ha$ luminosity, help us \Prasun{to} narrow \Prasun{down} the possible ranges of preshock velocity and number density. As a result, better \revisec{constraints} on mass accretion rate and protoplanet mass \Prasun{gives rise to} deeper insight into late-stage accretion \revisec{onto} gas giants.

At present, however, the $\Ha$ line profiles for the two protoplanets are resolved \revisec{with only} a few wavelength bins. 
Also, the 10\% and 50\% full widths, which we have used for constraining the ranges of $\MP$ and $\Mdot$, are estimated with Gaussian fitting \citep{Haffert+2019}. 
As demonstrated in section~\ref{sec:width}, not only emission but also absorption of $\Ha$ is crucial for constraining the mass accretion rate. Gaussian fitting may not be sufficient to represent
the \Prasun{absorbing} features. 
In addition, some theoretical and observational studies of proto-stellar accretion reported that for some targets, pre-shock gas absorbs a significant amount of the $\Ha$ radiation from 
\Prasun{the post-shock region,giving rise to} asymmetric spectral features \citep[e.g.,][]{Edward+1994}. 
\Prasun{On the other hand,} no such absorption occurs in the case of protoplanetary $\Ha$ emission \citep{Haffert+2019}. 
Higher-resolution spectroscopy is expected to cast light on protoplanetary gas accretion.
 
\acknowledgments
We thank the anonymous referee who helped us improve this paper greatly.
We are grateful to Gabriel-Dominique Marleau for the useful discussion. Also, we express our thanks to Prasun Dhang for significantly improving the manuscript.
This work is supported by JSPS KAKENHI Grant Numbers 17H01153 and 18H05439 and JSPS Core-to-Core Program ``International Network of Planetary Sciences (Planet$^2$)."

\begin{figure}
    \plotone{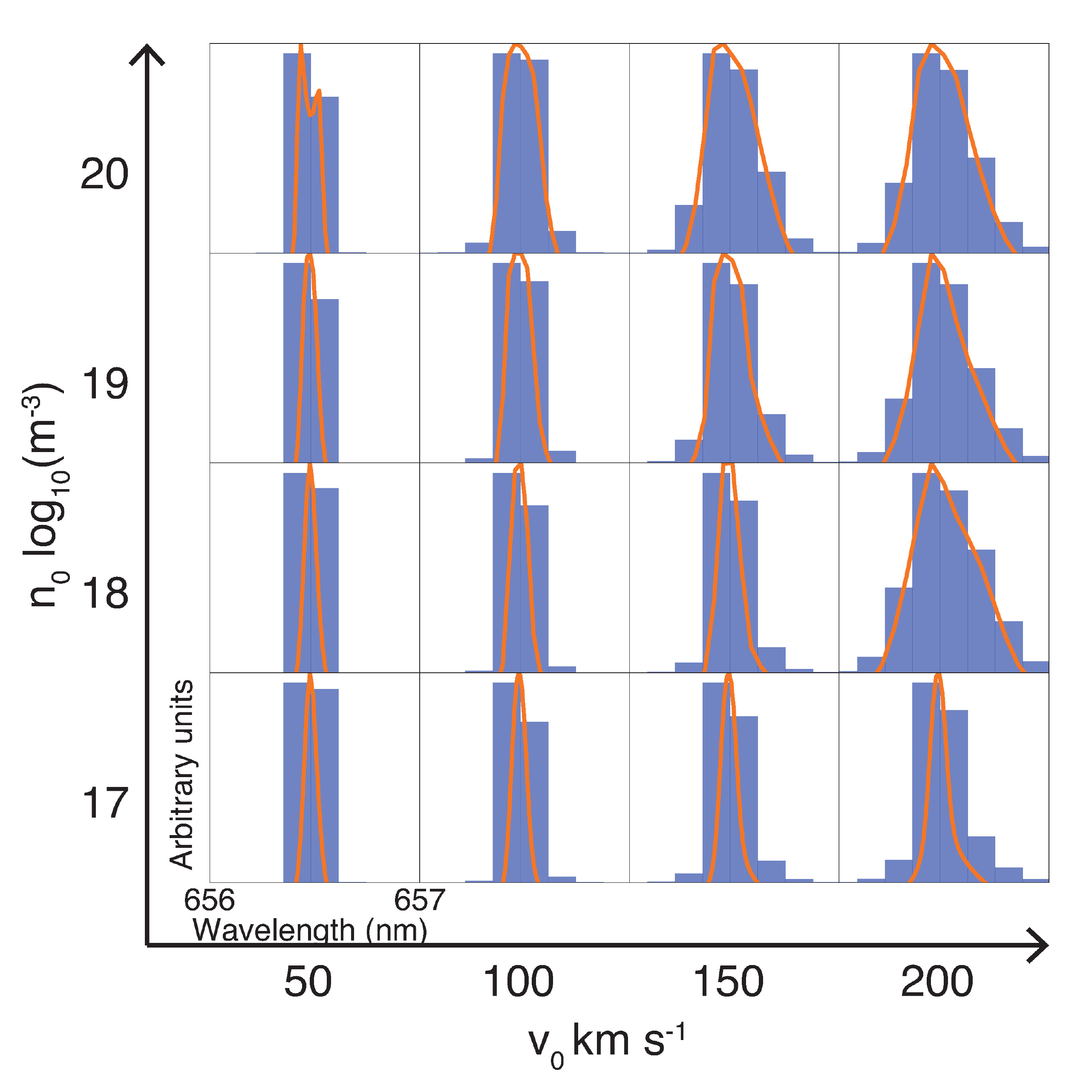}
    \caption{Simulated emission profiles around the H$\alpha$ line \reviseb{(orange line)} for several different choices of the \reviseb{preshock velocity,} $v_0=50$, $100$, $150$, and $200\,\mathrm{km\,s^{-1}}$ and \reviseb{the hydrogen number density at the shock,} $n_0= 1 \times 10^{17}$, $10^{18}$, $10^{19}$, and $10^{20}\,\mathrm{m}^{-3}$. Each panel is the same as Fig.~\ref{fig:profile} except for the values of $v_0$ and $n_0$.
    \reviseb{
    The background histogram indicates the simulated emission profiles smoothed over the bins of width 0.13~nm, which corresponds to \revisec{a} resolution of $R$ = 2500.}
    }
    \label{fig:multi}
\end{figure}

\appendix
We demonstrate the effects of \reviseb{the preshock velocity} $v_0$ and \reviseb{the hydrogen number density at the shock} $n_0$ on the $\Ha$ line profiles in \reviseb{Fig.}~\ref{fig:multi}.
The profile width becomes broader with increasing $v_0$, because larger $v_0$ results in hotter gas and thereby causes wider Doppler broadening after the shock. 
\reviseb{The sensitivity of the width to $v_0$ is, however, found not to be high. This is because H$\alpha$ emission mainly comes from deeper regions where the gas cools to $\sim10^4$\,K \citep[see][]{Aoyama+2018} and, therefore, the temperature of the emission region depends little on $v_0$.}
Note that \revisec{a} spectral resolution of $R=2500$ is not \reviseb{sufficient} to \revisec{distinguish} such a difference. 
\reviseb{
An increase in $n_0$ leads to broadening the profile width, because the contribution of the broad component coming from hot regions just after the shock becomes dominant, as $n_0$ becomes large (e.g., see the case of $v_0=200\,\mathrm{km\,s^{-1}}$ and $n_0=10^{18}\,\mathrm{m^{-3}}$).}
This broad component of the profile can be detected with \revisec{a} resolution \reviseb{of $R=2500$}.

\bibliography{list}

\begin{thebibliography}{}
\expandafter\ifx\csname natexlab\endcsname\relax\def\natexlab#1{#1}\fi
\providecommand{\url}[1]{\href{#1}{#1}}

\bibitem[{{Allen}(1976)}]{Allen3rd}
{Allen}, C.~W. 1976, {Astrophysical Quantities} (London: Athlone)

\bibitem[{{Aoyama} {et~al.}(2018){Aoyama}, {Ikoma}, \&
  {Tanigawa}}]{Aoyama+2018}
{Aoyama}, Y., {Ikoma}, M., \& {Tanigawa}, T. 2018, \apj, 866, 84

\bibitem[{{Aoyama} {et~al.}(in\,prep.){Aoyama}, {Marleau}, {Mordasini}, \&
  {Ikoma}}]{AGCI2019}
{Aoyama}, Y., {Marleau}, G.-D., {Mordasini}, C., \& {Ikoma}, M. in\,prep.

\bibitem[{{Batygin}(2018)}]{Batygin2018}
{Batygin}, K. 2018, \aj, 155, 178

\bibitem[{{Christensen} {et~al.}(2009){Christensen}, {Holzwarth}, \&
  {Reiners}}]{Christensen+2009}
{Christensen}, U.~R., {Holzwarth}, V., \& {Reiners}, A. 2009, \nat, 457, 167

\bibitem[{{Close} {et~al.}(2014){Close}, {Follette}, {Males}, {Puglisi},
  {Xompero}, {Apai}, {Najita}, {Weinberger}, {Morzinski}, {Rodigas}, {Hinz},
  {Bailey}, \& {Briguglio}}]{Close+2014}
{Close}, L.~M., {Follette}, K.~B., {Males}, J.~R., {et~al.} 2014, \apjl, 781,
  L30

\bibitem[{{Edwards} {et~al.}(1994){Edwards}, {Hartigan}, {Ghandour}, \&
  {Andrulis}}]{Edward+1994}
{Edwards}, S., {Hartigan}, P., {Ghandour}, L., \& {Andrulis}, C. 1994, \aj,
  108, 1056

\bibitem[{{Haffert} {et~al.}(2019){Haffert}, {Bohn}, {de Boer}, {Snellen},
  {Brinchmann}, {Girard}, {Keller}, \& {Bacon}}]{Haffert+2019}
{Haffert}, S.~Y., {Bohn}, A.~J., {de Boer}, J., {et~al.} 2019, Nature
  Astronomy, 329

\bibitem[{{Hartmann} {et~al.}(2016){Hartmann}, {Herczeg}, \&
  {Calvet}}]{Hartmann+2016}
{Hartmann}, L., {Herczeg}, G., \& {Calvet}, N. 2016, \araa, 54, 135

\bibitem[{{Hartmann} {et~al.}(1994){Hartmann}, {Hewett}, \&
  {Calvet}}]{Hartmann+1994}
{Hartmann}, L., {Hewett}, R., \& {Calvet}, N. 1994, \apj, 426, 669

\bibitem[{{Herczeg} \& {Hillenbrand}(2008)}]{Herczeg+Hillenbrand2008}
{Herczeg}, G.~J., \& {Hillenbrand}, L.~A. 2008, \apj, 681, 594

\bibitem[{{Koenigl}(1991)}]{Konigl1991}
{Koenigl}, A. 1991, \apjl, 370, L39

\bibitem[{{M{\"u}ller} {et~al.}(2018){M{\"u}ller}, {Keppler}, {Henning},
  {Samland}, {Chauvin}, {Beust}, {Maire}, {Molaverdikhani}, {van Boekel},
  {Benisty}, {Boccaletti}, {Bonnefoy}, {Cantalloube}, {Charnay}, {Baudino},
  {Gennaro}, {Long}, {Cheetham}, {Desidera}, {Feldt}, {Fusco}, {Girard},
  {Gratton}, {Hagelberg}, {Janson}, {Lagrange}, {Langlois}, {Lazzoni}, {Ligi},
  {M{\'e}nard}, {Mesa}, {Meyer}, {Molli{\`e}re}, {Mordasini}, {Moulin},
  {Pavlov}, {Pawellek}, {Quanz}, {Ramos}, {Rouan}, {Sissa}, {Stadler}, {Vigan},
  {Wahhaj}, {Weber}, \& {Zurlo}}]{Muller+2018}
{M{\"u}ller}, A., {Keppler}, M., {Henning}, T., {et~al.} 2018, \aap, 617, L2

\bibitem[{{Natta} {et~al.}(2004){Natta}, {Testi}, {Muzerolle}, {Randich},
  {Comer{￥'o}n}, \& {Persi}}]{Natta+2004}
{Natta}, A., {Testi}, L., {Muzerolle}, J., {et~al.} 2004, ￥aap, 424, 603

\bibitem[{{Rigliaco} {et~al.}(2012){Rigliaco}, {Natta}, {Testi}, {Randich},
  {Alcal{\`a}}, {Covino}, \& {Stelzer}}]{Rigliaco+2012}
{Rigliaco}, E., {Natta}, A., {Testi}, L., {et~al.} 2012, \aap, 548, A56

\bibitem[{{Sallum} {et~al.}(2015){Sallum}, {Follette}, {Eisner}, {Close},
  {Hinz}, {Kratter}, {Males}, {Skemer}, {Macintosh}, {Tuthill}, {Bailey},
  {Defr{\`e}re}, {Morzinski}, {Rodigas}, {Spalding}, {Vaz}, \&
  {Weinberger}}]{Sallum+2015}
{Sallum}, S., {Follette}, K.~B., {Eisner}, J.~A., {et~al.} 2015, \nat, 527, 342

\bibitem[{{Vriens} \& {Smeets}(1980)}]{Vriens+Smeets1980}
{Vriens}, L., \& {Smeets}, A.~H.~M. 1980, \pra, 22, 940

\bibitem[{{Wagner} {et~al.}(2018){Wagner}, {Follete}, {Close}, {Apai}, {Gibbs},
  {Keppler}, {M{\"u}ller}, {Henning}, {Kasper}, {Wu}, {Long}, {Males},
  {Morzinski}, \& {McClure}}]{Wagner+2018}
{Wagner}, K., {Follete}, K.~B., {Close}, L.~M., {et~al.} 2018, \apjl, 863, L8

\end{thebibliography}
\end{document}